# Indicator & crowding Distance-Based Evolutionary Algorithm for Combined Heat and Power Economic Emission Dispatch

Jiaze Sun [a,b], Jiahui Deng [a *], Yang Li [c]

[a] *School of Computer Science and Technology, Xi'an University of Posts and Telecommunications, Xi'an 710121, China*

[b] *Shaanxi Key Laboratory of Network Data Intelligent Processing, Xi'an University of Posts and Telecommunications, Xi'an 710121, China*

[c] *School of Electrical Engineering, Northeast Electric Power University, Jilin 132012, China*

## ABSTRACT

Heat and power have become the most indispensable resources. However, the traditional ways of generating power and heat are inefficient and cause high pollution; a CHP (Combined Heat and Power) unit can solve these problems well. In recent years, more attention has been paid to energy conservation and environmental protection, and Combined Heat and Power Economic Emission Dispatch (CHPEED) has become an important multi-objective optimization problem. In this paper, an Indicator & crowding Distance-based Evolutionary Algorithm (IDBEA) is put forward for handling this non-convex and non-linear problem. With consideration of the valve-point effects and power transmission loss, IDBEA is tested on three standard test systems with different types, including four units, five units and seven units. In the experiment, IDBEA is compared with several evolutionary algorithms, the simulation results demonstrate that IDBEA has strong stability and superiority, while the solutions show better convergence and diversity than several typical algorithms.

**NOMENCLATURE**

**Acronyms**

| | |
|---|---|
| CHP | combined heat and power |
| CHPED | CHP economic dispatch |
| CHPEED | CHP economic emission dispatch |
| BA | bat algorithm |

| | | |
|---|---|---|
| ABC | | artificial bee colony algorithm |
| TSCO | | social cognitive optimization algorithm with Tent map |
| ELD | | economic load dispatch |
| ADE-MMS | | a self-adaptable differential evolution algorithm integrating with multiple mutation strategies |
| DE | | differential evolution algorithm |
| NSGA-II | | non-dominated sorting genetic algorithm-II |
| GWO | | grey wolf optimization |
| TVA-PSO | | time-varying acceleration particle swarm optimization |
| MLCA | | multi-objective line-up competition algorithm |
| PF | | Pareto front |
| IDBEA | | indicator & crowding distance-based evolutionary algorithm |
| MOPs | | multi-objective optimization problems |
| FOR | | feasible operation region |
| POF | | Pareto-optimal front |
| IBEA | | indicator-based evolutionary algorithm |
| SPEA2 | | strength Pareto evolutionary algorithm 2 |
| PSO | | particle swarm optimization |
| PAES | | pareto archived evolution strategy |
| ABYSS | | archive-based hybrid scatter search |
| EAF | | the empirical attainment function |

**Symbols**

| | | |
|---|---|---|
| $P_i^{min}$, $P_i^{max}$ | | the lower and upper limit of the power output of the ith unit |
| $e_i, f_i$ | | the cost change coefficients of ith power-only unit |
| $C_{Total}$ | | total fuel cost |
| $C_{p,i}$ | | fuel costs of the ith power-only unit |
| $C_{c,j}$ | | fuel costs of jth CHP unit |
| $C_{h,k}$ | | fuel costs of the kth heat-only unit |
| $N_p$ | | number of power-only units |

| Symbol | Description |
|---|---|
| $N_c$ | number of CHP units |
| $N_h$ | number of heat-only units |
| $a_i$, $b_i$, $d_i$ | cost coefficients of ith power-only unit |
| $\alpha_j$, $\beta_j$, $\gamma_j$, $\delta_j$, $\varepsilon_j$, $\xi_j$ | cost coefficients of jth CHP unit |
| $\varphi_k$, $\eta_k$, $\lambda_k$ | cost coefficients of kth heat-only unit |
| $P_i$, $O_j$ | generating capacities of the ith power-only unit and jth CHP unit |
| $H_j$, $T_k$ | the heat generated by jth CHP unit and kth heat-only unit |
| $E_{Total}$ | total pollutant emissions |
| $E_S$ | total emission of $SO_2$ and $NO_x$ |
| $E_C$ | emission of $CO_2$ |
| $\mu_i$, $\kappa_i$, $\pi_i$, $\sigma_i$, $\nu_i$ | $SO_2$ and $NO_x$ emission factor of the ith power-only unit |
| $\tau_j$, $\rho_k$ | $SO_2$, $NO_x$ emission factors of jth CHP unit and kth heat-only unit |
| $\theta_i$ | $CO_2$ emission factors of ith power-only unit |
| $\psi_j$ | $CO_2$ emission factors of jth CHP unit |
| $\varpi_k$ | $CO_2$ emission factors of kth heat-only unit |
| $P_D$ | power demand |
| $P_L$ | power transmission loss |
| $B_{ij}$ | loss coefficient between the ith unit and jth unit |
| $B_{0i}$ | loss coefficient of the ith unit |
| $B_{00}$ | loss coefficient parameter |
| $H_D$ | total heat demand |
| $H_k^{min}$, $H_k^{max}$ | the lower and upper limit of the heat output of the kth unit |
| $P_j$ | the real power output of the jth unit |
| $H_j$ | real heat output of jth unit |
| x | decision vector |
| X | decision space |
| y | target vector |
| Y | target space |
| m | the number of decision variables |
| n | the number of objective functions |

| | | |
|---|---|---|
| f | | objective function |
| R | | number of equality constraints |
| L | | number of inequality constraints |
| $g_r$ | | equality constraints |
| $h_l$ | | inequality constraints |
| I | | binary quality indicator |
| a, b | | decision vectors |
| Z* | | reference point |
| k* | | scaling factor |
| i | | individual in the population |
| $P[i]_{dis}$ | | crowding-distance of ith individual |
| $P[i].n$ | | function value of ith individual of nth objective function |
| $f_n^{max}, f_n^{min}$ | | maximum and minimum function values of the nth objective function |
| s | | number of solutions |
| P | | initial population |
| A | | archive population |
| Q | | temporary mating pool |
| S | | offspring population |
| N | | size of population P, A, S |
| E | | evolutionary number |
| M | | generation counter |
| $x_j^L, x_j^U$ | | the lower and upper bounds of the individual x on the jth objective |

# 1 Introduction

## 1.1 Background information

Energy conservation has become the focus of global energy research [1] because of the rapid growth of energy demand and the resulting pollution increase. In fact, traditional thermal power plant does not produce the thermal energy into electrical energy efficiently during the power generation process. In the process, a large amount of thermal energy is wasted. The efficiency of energy conversion in the state-of-the-art combined cycle power plant is only 50% to 60% [2]. Combined Heat and Power (CHP) is a cogeneration unit, which is not only mature and reliable but also more flexible. CHP has higher efficiency and less effect on the environment than traditional thermal power technology [3]. CHP makes full use of the residual heat in converting fossil fuel directly into electrical energy, which increases the conversion rate and achieves 90% energy efficiency [4]. Compared with traditional pure electric generating units and pure heat generating units, cogeneration units can save 10% to 40% of the generation cost, which means that less fuel is consumed for the same amount of heat and electricity [5]. In addition, as an environmentally friendly system, cogeneration units can reduce greenhouse gas emissions by 13% to 18% [6]in comparison with the conventional generating units. Combined Heat and Power Economic Dispatch (CHPED) is to optimize the distribution of heat load and power load commands to reduce fuel costs [7]. However, with the increasing prominence of social environmental problems, economic optimization alone cannot meet the needs of social energy conservation and environmental protection. Thus, the Combined Heat and Power Economic Emission Dispatch (CHPEED) is a typical multi-objective optimization problem under a series of equality and inequality constraints. Economic emission dispatch has two conflicting objectives: pollution emission minimization and fuel cost minimum. The CHPEED problem is to search a set of optimal feasible solutions while achieving the minimum fuel costs and pollutant emissions. Moreover, The CHPEED problem is always non-linear, non-convex, non-smooth and multi-constrained.

Therefore, searching for optimal solutions is a challenging problem [8].

## 1.2 Literature survey

CHPED and CHPEED problems have been extensively studied, the methods for solving them are roughly divided into two types: traditional mathematical methods and heuristic algorithms. Traditional mathematical methods usually solve the CHPED problem by Lagrange multiplier, linear programming, quadratic programming or dynamic programming [9] [10]. However, as the cost curve model of the CHPEED problem is highly non-linear, non-monotonous and sometimes discontinuous, traditional mathematical methods are not applicable. Literature [11] integrates traditional mathematical methods and heuristic algorithms to solve multi-objective problems. As a simple and efficient intelligent optimization algorithm, the heuristic algorithm has achieved great success in solving nonlinear problems in the fields of production scheduling, system control, pattern recognition, artificial intelligence, computer engineering and so on. The heuristic algorithm can be used to solve combinatorial optimization problems and numerical optimization problems. It has been widely accepted as an efficient optimization method in the engineering field, which is more suitable for solving non-linear, non-convex and non-smooth constraints.

Literature [4] [12] uses Cuckoo search algorithm and Group search optimization algorithm to solve the CHPED problem. Literature [13] proposes an integrated technique based on CSO with PPS method to solve the CHPED problem, while literature [14] solves this problem by fusing Bat Algorithm (BA) and Artificial Bee Colony (ABC) algorithm. TSCO algorithm is proposed in the literature [15] to solve this problem. Economic load dispatch (ELD) is also an important issue in single-objective problem, literature [16] proposes a self-adaptable differential evolution algorithm integrating with multiple mutation strategies (ADE-MMS) for this problem, the method extends the differential evolution algorithm (DE) and the experimental results show great potential to solve the ELD problems.

With the increasing environmental protection awareness, the study on CHPEED has been carried out extensively. At present, three kinds of multi-objective evolutionary algorithm paradigms are widely used [17]: Pareto-based [18] approaches, indicator-based approaches and decomposition-based [19] approaches. The most classical Pareto-based algorithm is NSGA-II [20]. Basu M proposed NSGA-II [21] to solve the CHPEED problem. But the poor convergence hinders the wide application of the algorithm. Multi-Objective Particle Swarm Optimization (MOPSO) [22] algorithm is used to handle this problem. Time-Varying Accelerated Particle Swarm Optimization (TVA-PSO) [23] algorithm is proposed to deal with the problem. The Multi-objective Line-up Competition Algorithm (MLCA) is submitted to solve the problem by extracting compromise solutions from Pareto optimal solutions using fuzzy decision [24]. A two-stage algorithm is presented to provide different solutions according to different decision makers [7]. In a word, the algorithms mentioned in the above literature are the Pareto-based evolutionary algorithms, while the effectiveness of the algorithms depends on the shape and distribution of the Pareto Front (PF) of the problem. The decomposition-based multi-objective evolutionary algorithm [25] with complex PF cannot guarantee that the obtained solutions are uniformly distributed in Pareto optimum when the weight vector with uniform distribution, so the algorithm cannot obtain a better solution. Indicator-Based Evolutionary Algorithm (IBEA) [26] assigns different fitness values to every group of solutions according to different problem needs, IBEA has well-convergent, the nevertheless, its disadvantage is the poor diversity of solutions, and the selection of indicator is the key point. Regarding the CHPEED problem, incorporating various methods may bring new ideas.

### 1.3    The work of this paper

In this paper, a model of economic emission dispatch for CHP is constructed. The two objectives of fuel cost and pollutant emissions are optimized simultaneously, and the constraints of the problem are formulated. In CHPEED modeling, the impacts of valve-point effects and transmission loss are considered, which will happen in the real dispatching situation. The occurrence of valve-point effects may arise in power losses, and the transmission loss between units will lead to the reduction of power generation. This paper puts forward a new algorithm based on IBEA, namely Indicator & crowding Distance-Based Evolutionary Algorithm (IDBEA). IDBEA added a crowding-distance operator for IBEA to calculate the distances of the solutions. IDBEA not only maintain the convergence of IBEA very well but also can increase the diversity of solutions effectively by the crowding-distance operator. Therefore, the solution set is closest to the real PF and the optimal scheduling scheme can be obtained. To validate the effectiveness and superiority of IDBEA, several different types of standard test systems are used, while the results are compared with other typical algorithms. For the sake of verifying the universality and applicability of IDBEA in both single-objective and multi-objective problems, thus the algorithm is tested on CHP economic dispatch problem and CHP economic emission dispatch problems.

The main contributions of this paper:

1. In this paper, an Indicator & crowding Distance-Based Evolutionary Algorithm (IDBEA) has been put forward to resolve the CHPEED problem.

2. Both valve-point effects and transmission loss are considered in the CHPEED model. And environmental protection is paid more attention to solving.

3. In the simulation experiments, CHPED and CHPEED problems are used to validate the proposed algorithm on three standard test systems respectively, which prove good universality and superiority of the algorithm.

### 1.4 Organization of this paper

The remaining sections of this article are structured as follows. Section 2 formulates the objectives and constraints of the CHPEED problem. Section 3 describes the Multi-objective Optimization Problem (MOP). Section 4 describes the proposed algorithm in this paper in detail and provides the flowchart of the algorithm. Section 5 carries out experiments on three test systems and compares IDBEA with some typical algorithms. Section 6 is a summary of this article.

## 2 Formulation of CHPEED problem

The CHPEED problem is to search for an optimal set of solutions that obtained the minimum fuel costs and the minimum pollutant emissions simultaneously. In the process of searching, we need to satisfy the equality constraints and inequality constraints. The test systems used in this paper include three types of units: power-only, cogeneration and heat-only. In the systems, power-only units and cogeneration units generate power output, while heat is produced by cogeneration units and heat-only units.

### 2.1 Objective functions

In the CHPEED problem, the minimum fuel costs and the minimum pollutant emissions are hoped to be obtained at the same time, hence the objective function of the CHPEED problem includes fuel costs and gas emission.

### 2.1.1 Fuel costs

Valve-point effects occurs in the work of power-only units. It refers to the phenomenon that when the valve suddenly opens, steam loss leads to increased consumption and superimposes unit consumption curve and pulsation effect [5], which is described in Fig.1 [27]:

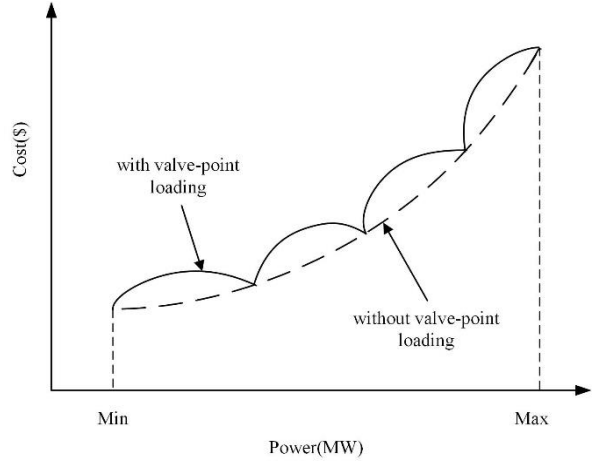

**Fig 1 Illustration of the valve-point effects.**

As shown in Fig.1, Min and Max are the lower upper limits of the power-only unit capacity. The change function of consumption characteristics of traditional units caused by valve-point effects is $\left|e_i \sin\{f_i(P_i^{min} - P_i)\}\right|$, where $P_i$ represents the real power output of the ith unit, $P_i^{min}$ represents the lower limit of the power output of the ith unit, $e_i$ and $f_i$ are the cost change coefficients of the ith power-only unit. The total production cost can be described as the following formula (1):

$$C_{Total} = \sum_{i=1}^{N_p} C_{p,i}(P_i) + \sum_{j=1}^{N_c} C_{c,j}(O_j, H_j) + \sum_{k=1}^{N_h} C_{h,k}(T_k)$$

$$= \sum_{i=1}^{N_p}[a_i + b_i P_i + d_i P_i^2 + |e_i \sin\{f_i(P_i^{min} - P_i)\}|]$$

$$+ \sum_{j=1}^{N_c}[\alpha_j + \beta_j O_j + \gamma_j O_j^2 + \delta_j H_j + \varepsilon_j H_j^2 + \xi_j O_j H_j]$$

$$+ \sum_{k=1}^{N_h}[\varphi_k + \eta_k T_k + \lambda_k T_k^2] \qquad (1)$$

Where, $C_{Total}$ represents the total fuel cost; $C_{p,i}$, $C_{c,j}$, $C_{h,k}$ represent the respective fuel costs of the ith power-only unit, jth CHP unit, and kth heat-only unit; $N_p$, $N_c$, $N_h$ represent the respective numbers of power-only units, CHP units, and heat-only units; $a_i$, $b_i$, $d_i$ represent the cost coefficients of the ith power-only unit; $\alpha_j$, $\beta_j$, $\gamma_j$, $\delta_j$, $\varepsilon_j$, $\xi_j$ represent the cost coefficients of the jth CHP unit; $\varphi_k$, $\eta_k$, $\lambda_k$ represent the cost coefficients of the kth heat-only unit; $P_i$, $O_j$ represent the respective generating capacities of the ith power-only unit and jth CHP unit; $H_j$, $T_k$ represent the heat generated by the jth CHP unit and the kth heat-only unit, respectively.

### 2.1.2 Gas emission

When the units work, they produce polluted gases, which mainly include $SO_2$, $NO_x$ and $CO_2$. The emission of gases can be described as follows:

$$E_{Total} = E_S + E_C \tag{2}$$

$$E_S = \sum_{i=1}^{N_p} E_{p,i}(P_i) + \sum_{j=1}^{N_c} E_{c,j}(O_j) + \sum_{k=1}^{N_h} E_{h,k}(T_k)$$

$$= \sum_{i=1}^{N_p}[\mu_i + \kappa_i P_i + \pi_i P_i^2 + \sigma_i e^{(\nu_i P_i)}] + \sum_{j=1}^{N_c} \tau_j O_j + \sum_{k=1}^{N_h} \rho_k T_k \tag{3}$$

$$E_C = \sum_{i=1}^{N_p} \theta_i P_i + \sum_{j=1}^{N_c} \psi_j P_j + \sum_{k=1}^{N_h} \varpi_k P_k \tag{4}$$

Where, $E_{Total}$ represents the total pollutant emissions, $E_S$ represents the total emission of $SO_2$ and $NO_x$, while $E_C$ represents the emission of $CO_2$; $\mu_i$, $\kappa_i$, $\pi_i$, $\sigma_i$, $\nu_i$ represent the $SO_2$ and $NO_x$ emission factor of the ith power-only unit; $\tau_j$ and $\rho_k$ represent the $SO_2$, $NO_x$ emission factors of the jth CHP unit and the kth heat-only unit, respectively ; $\theta_i$, $\psi_j$, $\varpi_k$ represent the respective $CO_2$ emission factors of the ith, jth, kth different types of unit, respectively.

## 2.2 Constraints

In the CHP system, first of all, the balance constraints of heat and power should be satisfied. Secondly, the units in the CHP system should meet different upper and lower constraints. Hence, there have equality constraints and inequality constraints.

### 2.2.1 Equality constraints

The mathematical expression of power transmission loss between units is given by formula (5):

$$P_L = \sum_{i=1}^{N_p}\sum_{j=1}^{N_p} P_i B_{ij} P_j + \sum_{i=1}^{N_p}\sum_{j=1}^{N_c} P_i B_{ij} O_j + \sum_{i=1}^{N_c}\sum_{j=1}^{N_c} O_i B_{ij} O_j$$

$$+ \sum_{i=1}^{N_p} B_{0i} P_i + \sum_{i=1}^{N_c} B_{0i} O_i + B_{00} \tag{5}$$

Where, $B_{ij}$ represents the loss coefficient between the ith unit and the jth unit, $B_{0i}$ represents the loss coefficient of the ith unit, $B_{00}$ represents the loss coefficient parameter.

Therefore, the power demand constraint is:

$$\sum_{i=1}^{N_p} P_i + \sum_{j=1}^{N_c} O_j = P_D + P_L \tag{6}$$

Where, $P_D$ and $P_L$ represent power demand given by system and power transmission loss, respectively.

The heat demand for the system is:

$$\sum_{j=1}^{N_c} H_i + \sum_{k=1}^{N_h} T_k = H_D \tag{7}$$

Where $H_D$ represents the total heat demand of the system.

### 2.2.2 Inequality constraints

Each unit has a specific output capacity limit, it's easy for power-only units and heat-only units to know their capacity limitations. But for cogeneration units, their power and heat output are mutually constrained, we need to describe the feasible operating region to analysis the capacity constraints. Fig.2 depicts the power feasible operating area [21].

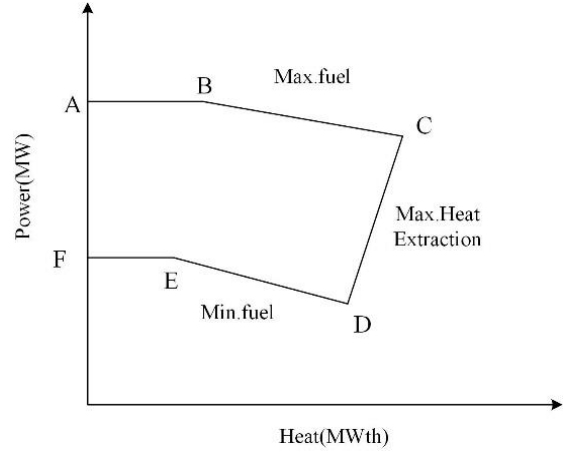

**Fig 2 Heat-power feasible operating region for cogeneration unit.**

The points A, B,…, F in Fig.2 are the feasible region coordinates [5], which means the capacity limitations of the cogeneration unit. The feasible area of the cogeneration unit is a closed area surrounded by ABCDEF. The relationship between heat output and power output can be seen from Fig.2. For example, along BC, the power output of the cogeneration unit decreases as heat output increases. While along DC, the power output of the cogeneration unit increases with the increases of heat output. So, the upper and lower power output limits we can get for the cogeneration unit are B and D, respectively. While the upper and lower heat output limits are F and C. On the basis of the above analysis, we can establish the inequality constraints of the units' capacity.

The following equations are capacity constraints for all types of units [7] [21]:

$$P_i^{min} \leq P_i \leq P_i^{max}, i = 1, \dots, N_p \tag{8}$$

$$P_j^{min}(H_j) \leq P_j \leq P_j^{max}(H_j), j = 1, \dots, N_c \tag{9}$$

$$H_j^{min}(P_j) \leq H_j \leq H_j^{max}(P_j), j = 1, \ldots, N_c \quad (10)$$

$$H_k^{min} \leq H_k \leq H_k^{max}, k = 1, \ldots, N_h \quad (11)$$

Equation (8) represents the power constraints of the power-only units. $P_i^{min}$ and $P_i^{max}$ represent the lower and upper limits of the power output of the ith unit, respectively. Equation (11) represents the heat constraints of the heat-only units. $H_k^{min}$ and $H_k^{max}$ represent the lower and upper limits of the heat output of the kth unit, respectively. The capacity limits of the power-only and heat-only units can be found in [5]. Equation (9) and equation (10) represent the power and heat constraints of the cogeneration units, respectively, $P_j$ and $H_j$ represent the real power and heat output of the jth unit, respectively. As described in Fig.2, we can get the minimum and maximum capacity of power and heat output.

## 3   Principle of multi-objective optimization

There are many complex Multi-objective Optimization Problems (MOPs) in factual engineering. As described in the second section, the CHPEED problem is also a typical MOP, which optimizes two objective functions at the same time. MOPs usually have a series of constraints, including equality constraints and inequality constraints. It can be described as the mapping of m decision vectors x on n objective functions f under the condition that R equality constraints and L inequality constraints are satisfied. MOPs can be expressed as follows:

$$\min/\max y = f(x) = (f_1(x), f_2(x), \ldots, f_n(x)) \quad (12)$$

$$\text{subject to} \quad g_r(x) = 0 \quad r = 1, \ldots, R$$

$$h_l(x) \leq 0 \quad l = 1, \ldots, L$$

$$x = (x_1, x_2, \ldots, x_m) \in X$$

$$y = (y_1, y_2, \ldots, y_n) \in Y \quad (13)$$

In equation (12) and equation (13), x represents the decision vector, X represents the decision space, y represents the target vector, and Y represents the target space. $g_r$ and $h_l$ represent equality constraints and inequality constraints, respectively.

Generally speaking, the objective functions of MOPs are contradictory. Therefore, the MOPs solution is a set of optimal solutions, in which there is no better solution than other solutions, called Pareto optimal solution set. The set of optimal solution sets mapped onto the objective function is called the Pareto Front (PF). The mathematical concept of Pareto optimal is described as follows [28]: assume this is a minimization problem, there are two decision vectors $a, b \in X$, if a dominates b (denoted as $a \prec b$), if and only if the following conditions are met:

$$\forall i \in \{1,2,\ldots,n\}: f_i(a) \leq f_i(b) \quad \wedge$$

$$\exists j \in \{1,2,\ldots,n\}: f_j(a) < f_j(b). \quad (14)$$

That is to say, for decision vectors a and b, the function value of a is not greater than that of b on any objective function, and at least there has one objective function that the function value of a is smaller than the function value of b, which is called $a \prec b$.

# 4 Indicator & crowding Distance-Based Evolutionary Algorithm

IBEA is a kind of optimization algorithm, which is used to solve optimization problems. IBEA is an algorithm based on the Hypervolume indicator. There are a few indicators can be used to evaluate whether the solutions we obtained are converged or widely distributed to the real PF. Among them, the Hypervolume indicator can evaluate these two performances comprehensively. Generally, the evaluation of the ability of evolutionary algorithms to solve problems can be divided into two aspects: convergence and diversity. While IBEA has the issues of diversity in searching solutions [29]. In order to improve the convergence and diversity of the algorithms for the CHPEED problem, we proposed an Indicator & crowding Distance-Based Evolutionary Algorithm (IDBEA). On the basis of the Hypervolume indicator, crowding-distance is introduced IDBEA to select individuals with larger distance values to ensure the diversity of the population. Therefore, the new algorithm not only guarantees the convergence of the solutions but also increases the diversity of the solutions. Two key ideas in the new algorithm will be described in detail in sections 4.1 and 4.2, the algorithm flow will be introduced in 4.3.

## 4.1 Hypervolume indicator

The quality indicator is a function that uses some preference information to assign a real number to any one of a number of approximate solution sets. In this way, the relative merits of any two approximate solution sets can be judged according to the real numbers corresponding to each approximate solution set. The indicator-based evolutionary algorithm [26] uses a unique fitness assignment method to remove the worst individuals with the largest fitness values in iterations, we will elaborate the worse point selection in section 4.3. This kind of algorithm has well-convergence and can be used to solve the problem with high target dimensions [30]. In the paper, IDBEA uses the Hypervolume indicator. The Hypervolume indicator is a comprehensive evaluation indicator that can assess convergence and diversity simultaneously. The evaluation results reflect the convergence of the algorithm, as well as the uniformity and breadth of the different solution sets, that is, the diversity of the solutions. Hypervolume of the solution set $S = (x_1, \ldots, x_n)$ is the volume of the region that is dominated by S in the objective space [29]. The Hypervolume indicator evaluation obeys the Pareto rule [31]. By calculating the volume of hypercubes surrounded by all points in the non-dominated solution set, and by the reference points in the target space, the Hypervolume indicator value of the solution set can be obtained. The larger the Hypervolume value, the better the overall performance of the algorithm.

The population P represents a sample of the decision space, and the fitness value of each individual represents the usefulness regarding the optimization goal. According to the fitness values, individuals are graded and sorted to selectively remove the worst individuals in the iterations. The formula for calculating fitness by using indicator is as follows (15):

$$F(x^1) = \sum_{x^2 \in P \setminus \{x^1\}} -e^{-I(\{x^2\},\{x^1\})/k^*} \tag{15}$$

Where $I$ is a binary quality indicator, and it represents the comparison of the quality of two Pareto set approximations relative to each other [26]. $k^*$ is a scaling factor that is greater than 0. The value of $k^*$ varies with the actual problem. In the CHPEED problem, the empirical proof that the algorithm has a relatively good operation result when $k^*$ is 0.05.

In this paper, the $I_{HD}$- indicator is used, which is based on the Hypervolume concept:

$$I_{HD}(A,B) = \begin{cases} I_H(B) - I_H(A) & if\ \forall x^2 \in B\ \exists x^1 \in A:\ x^1 \succ x^2 \\ I_H(A+B) - I_H(A) & else \end{cases} \quad (16)$$

In the above formula, $I_H(A)$ provides the Hypervolume of the objective space dominated by A, while $I_{HD}(A,B)$ means the volume of the space that is dominated by B but not by A with respect to a predefined reference point $Z^*$. We will explain it in detail in Fig.3 below [26], and A and B both contain one decision vector.

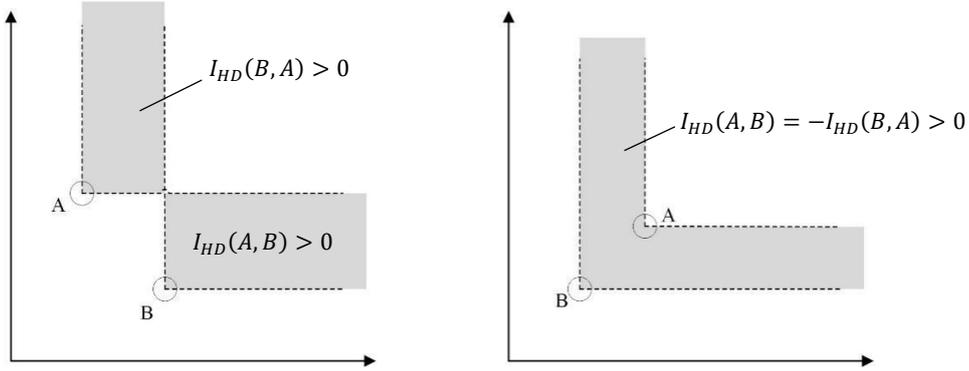

**Fig 3 Illustration of a binary quality indicator.**

### 4.2 Crowding-distance

To resolve the problem of poor diversity of the solutions in IBEA, the crowding-distance ranking strategy [20] is introduced to select individuals in the population and improve the performance of the algorithm. To obtain the crowding-distance of individuals in the population, we calculate the average distance values between two adjacent individuals for each target individual. As shown in Fig.4, the average distance of individual i is the average side length of the rectangle composed of $i-1$ and $i+1$ vertices. Points marked in filled circles are solutions of the same non-dominated front.

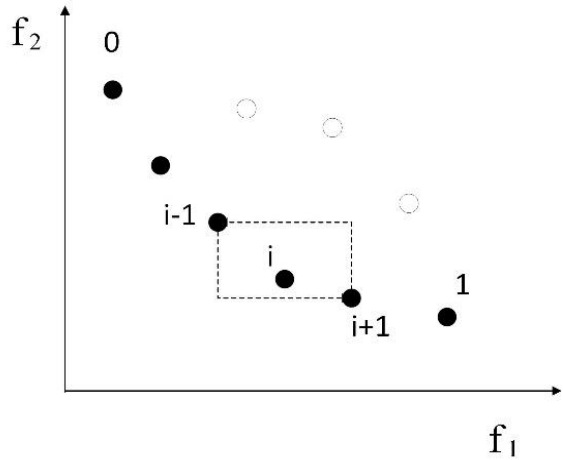

**Fig 4 Crowding-distance calculation.**

The initial crowding-distance of all individuals in the population has been set to 0.0 before the crowding-distance is calculated. Then, the individuals in the population are sorted by each objective function value in ascending order. Positive infinity distance value is assigned to each boundary individual, which is the maximum individual or minimum individual in the sorting. Meanwhile, the distance values of the remaining individuals are set to the absolute normalized difference in the function of the two adjacent individuals. Crowding-distance ranking strategy is also used in NSGA-II [20]. Algorithm 1 calculates the crowding-distance of solutions in non-dominated solution set P:

**Algorithm 1 crowding-distance calculation**

**Input: non-dominated solution set P, number of solutions s**

**Output: crowding-distance of each solution $P[i]_{dis}$**

1: for each i ∈ s, set $P[i]_{dis} = 0.0$ do

2: for each objective n do

3:    $P = sort(P, n)$

4:    $P[1]_{dis} = P[s]_{dis} = \infty$

5:    for $i = 2$ to $s - 1$ do

6:       $P[i]_{dis} = P[i]_{dis} + (P[i+1].n - P[i-1].n)/(f_n^{max} - f_n^{min})$

For the individual i, $P[i]_{dis}$ represents crowding-distance, while $P[i].n$ represents the function value of the nth objective function. $f_n^{max}$ and $f_n^{min}$ represent the maximum and minimum function values of the nth objective function, respectively. We can use Algorithm 1 to calculate the crowding-distance of all the solutions in the non-dominated solution set P. In each objective, each individual is sorted and crowding-distance is calculated. When the total crowding-distance is equal to the sum of the distance values of the individuals corresponding to each objective. The solutions with smaller crowding-distance are then removed to increase the diversity of the solutions. This is the key improvement of the original algorithm in this paper.

### 4.3 IDBEA

Since the IDBEA we proposed is based on IBEA, the main framework is the same as IBEA, the performance of the algorithm with Crowding-distance ranking strategy is improved by step (8). As described in sections 4.1 and 4.2, the detailed steps of the IDBEA are shown below:

Step (1). Initialization: Create a parent population P by generating random values in the feasible operating region. The feasible operating region in this work can be defined by $(rand(0,1) * (x_j^U - x_j^L) + x_j^L)$, where $x_j^U$ and $x_j^L$ represent the upper and lower bounds of the individual x in the jth objective, respectively. The bounds we used in this work are borrowed from Reference [5]. Create an archive population A and a temporary mating pool Q. Set the value of population size of P and A to N and the size of population Q to 2*N. Create an offspring population S with population size being N and with no individual in it. The number of maximum iterations is set to E, and the generation counter is $M = 0$.

Step (2). Fitness assignment: Combine individuals in parent population P with individuals in archive population A to Q (when the generation counter M = 0, population A keeps empty, Q equals P), then the fitness values of individuals in Q are calculated as equation (15).

Step (3). Environmental selection: Iterate through the following three steps, while the size of population Q is no more than N:

1. Rank the individuals in population Q by their fitness values.

2. The fitness values in the population Q are compared one by one as follows. If the fitness value of individual $a$ is greater than the fitness value of individual $b$ ($F(a) > F(b)$), then we will set individual $a$ to the individual with the largest fitness value. Otherwise, the algorithm will not change the worst individual. Until the end of all comparisons, we select the individual $x$ with the largest fitness value in population Q after sorting. If there is more than one individual with the same fitness value after sorting, we will select the first of them to have this fitness value.

3. Remove the individual $x$ from population Q, recalculate the fitness values of the remaining individuals in population Q. When the size of Q equals N, the individuals are copied into A.

Step (4). Termination: If $M \geq E$:

1. The iteration is stopped.

2. Rank population A with a non-dominated sorting algorithm and obtain the first non-dominated front.

3. The first non-dominated front in population A is outputted as the Pareto approximate solution set.

Otherwise, proceed with the following steps.

Step (5). Mating Selection: Using binary tournament selection method for individuals in archive population A to select two parent individuals.

Step (6). Crossover and Mutation: The simulated binary crossover has been used to do crossover in the parent individuals to generate the offspring individuals, then polynomial mutation is used for offspring individuals to keep the diversity of the population and prevent the population from falling into local optimum. The offspring individuals we have obtained are put into population S. When the size of population S is smaller than N, return to Step (5).

Step (7). Replace: Make the offspring individuals become the new parent individuals, which means replace parent population P with offspring population S; increment the counter: $M = M + 1$.

Step (8). Crowding-distance assignment and Ranking: Distribution of crowding-distance among individuals in population A and sorting by distance from large to small; retention of the first $N * 4/5$ individuals with larger distances in archive population A. Turn to Step (2).

The process of population setting and execution is shown in Fig. 5:

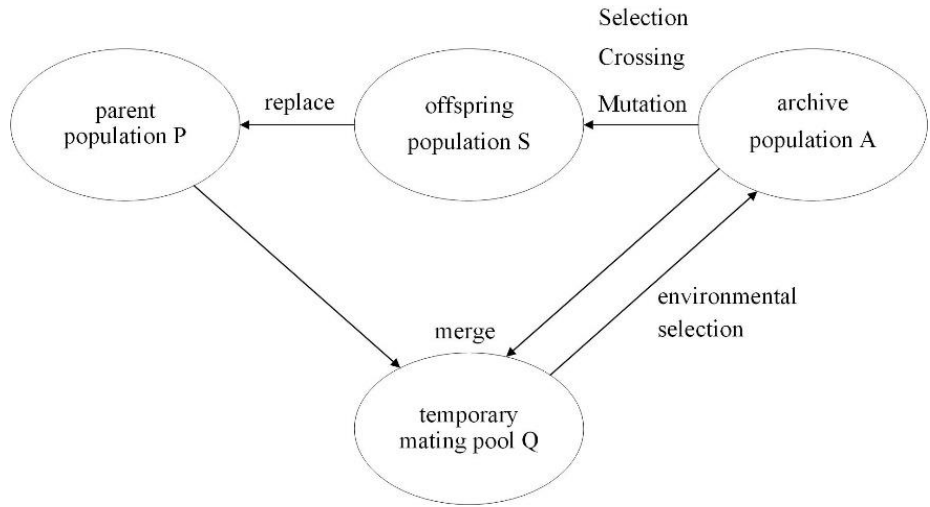

**Fig 5 Population execution process.**

According to the flow of IDBEA, parent population P is combined with archive population A, in temporary mating pool Q, the fitness value of each solution are assigned and sorted, the first N individuals are moved to population A, then do selection, crossing, and mutation, the parent population P is replaced with offspring population S, then continue the next iteration.

The computational flow chart of IDBEA is shown in Fig.6:

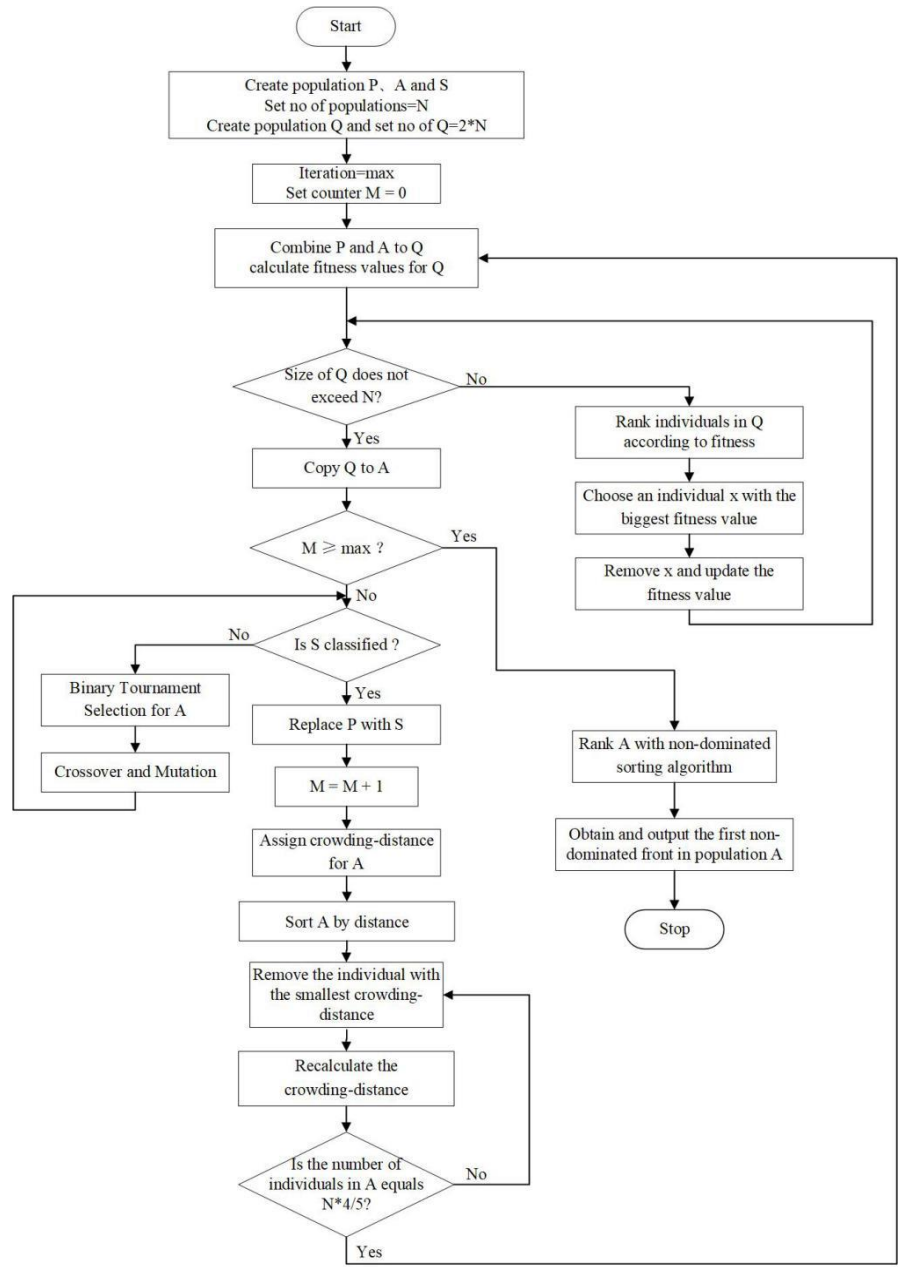

**Fig 6 Flowchart of the IDBEA.**

The Hypervolume indicator represents the natural extension of Pareto dominance, if there are two objectives, HV refers to the area enclosed between the optimal approximate front and the reference point (after normalization, the reference point of the minimization problem is (0,0) and the reference point of the maximization problem is (1.1,1.1)), and the larger the area is (the larger the indicator value is), the stronger the dominance relationship is. If there are three objectives, HV represents the volume of the optimal approximate front and the reference point. If there are more than three objectives, HV represents the hypervolume enclosed by the optimal approximate front and the reference point. While in single-objective problems, the partial order relationship of multi-dimensional objectives "degenerates" into the well-ordered relationship of a single objective. When we solve the single objective minimization problem, the individual with a smaller value "dominates" the individual with a larger value, HV represents the distance between the optimal approximate solution and the reference point (0, 0). In the process of fitness allocation, we actually calculate the numerical value of the objective function. The larger the HV value is, the more obvious the "dominance relationship" is. Finally, we will get the best approximate solution. In this way, we can calculate the fitness values of the individuals and evaluate the best approximate solution.

## 5 Simulation results

In this section, the results of IDBEA for solving CHPED and CHPEED problems are analyzed in detail. In CHPED, the single-objective optimization problem is used to test the universality of the algorithm, in which IDBEA degenerates into a single-objective algorithm. IDBEA is used on three standard test systems, including the single-objective test system for 4 units, and the multi-objective test systems for 5 units and 7 units. The algorithms for comparison have been implemented on multi-objective optimization framework jMetal 4.5. In all the algorithms we compared in this paper, the population size is 200, and the maximum iteration number is 25000, while crossover and mutation probabilities have been selected as 0.9 and 1/m, respectively, where m means the number of decision variables. All codes are implemented in Java language with JDK 1.8 and run on a PC platform with 2.8 GHz Intel Core i7 and 8 GB RAM. To analyze the quality of the simulation results objectively, the classical algorithms are used to compare with the IDBEA.

### 5.1 Test system 1

Test system 1 contains a power-only unit (unit 1), a heat-only unit (unit 4) and two cogeneration units (unit2-3), which is put forward by Gou et al. in [9]. This test system aims at evaluating the algorithm for the CHPED problem. So as to compare with the results in the literature, the influence of transmission loss and valve-point effects are ignored. In this test system, the power and heat demand are 200MW and 115MWth, respectively.

The best-found solution of IDBEA is shown in Table 1, where all the algorithms are run 30 times, the best-found solution of IDBEA is compared with those of the other typical algorithms, and the solutions of the other algorithms are also the best-found solutions.

Table 1 The best-found solution of IDBEA and other methods (Pd=200MW, Hd=115MWth).

| Methods | P1 | P2 | P3 | H2 | H3 | H4 | Cost($) | Time(s) |
|---|---|---|---|---|---|---|---|---|
| **PSO** | 0 | 160 | 40 | 40 | 75 | 0 | 9257.1 | 2.1 |
| **NSGA-II** | 0 | 159.9 | 40.1 | 40 | 75 | 0 | 9258.3 | 0.8 |
| **HS** | 0 | 160 | 40 | 40 | 75 | 0 | 9257.1 | 1.2 |
| **DE** | 0 | 159.9 | 40 | 40.1 | 75.1 | 0 | 9259.4 | 0.5 |
| **GWO** | 0 | 160 | 40 | 40 | 75 | 0 | 9257.1 | 0.9 |
| PAES | 0 | 152.5 | 33.7 | 47.5 | 81.3 | 0 | 9326.9 | 0.4 |
| ABYSS | 0.1 | 138 | 34.9 | 61.9 | 80.3 | 0 | 9568.9 | 0.5 |
| SPEA2 | 0 | 159.8 | 40.2 | 40 | 75 | 0 | 9259.7 | 1.4 |
| IBEA | 0 | 126.9 | 40 | 43 | 75 | 0 | 8439.5[a] | 2.8 |
| IDBEA | 0 | 159.5 | 40.5 | 39.8 | 75.3 | 0 | 9255.7 | 0.3 |

[a] Not feasible.

From the Feasible Operation Region (FOR) of CHP unit 2, the minimum total power generation of the units is 40MW, while PAES and ABYSS do not have actual power output. Therefore, the calculation results of PAES and ABYSS do not participate in the comparison. The minimum cost of IBEA is $8439.47, nevertheless, the real power produced an error of 33.14 MW. IBEA doesn't meet the power demand. The minimum cost of IDBEA is $9255.65 when it satisfies the heat and power requirements in the test system. In Table 1, the cost of IDBEA is reduced by $1.4, $2.6, $1.4, $3.7, $1.4 and $4 than PSO, NSGA-II, HS, DE, GWO, and SPEA2, respectively. Moreover, the worst cost of IDBEA among 30 runs is $9426.65, the average cost is $9311.4, the standard deviation of the results is 55.2. According to Table 1, IDBEA has less execution time than other algorithms under the same conditions.

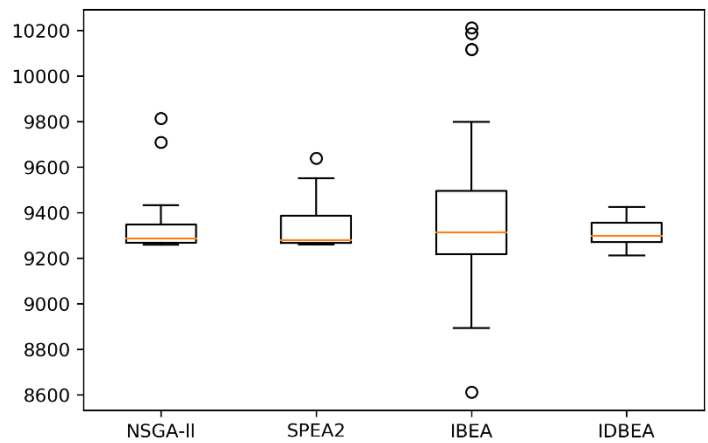

**Fig 7 The cost box plot of NSGA-II、SPEA2、IBEA and IDBEA for test system 1.**

Fig.7 is the box plot of the best-found solutions obtained by NSGA-II, SPEA2, IBEA, and IDBEA. Each algorithm runs 30 times and gets 30 best-found solutions, as is a statistical result, the best-found solutions can be used to illustrate the overall distribution of them, and the stability of the algorithms. From Fig.7 we can see that all three algorithms except IDBEA have obvious outliers, among which IBEA has more outliers and the solution distribution is the most dispersed. The solution distribution of IDBEA is more centralized and there are no outliers obviously. This shows that IDBEA has higher stability and better performance in searching for an effective solution than the other three algorithms in solving the CHPED problem.

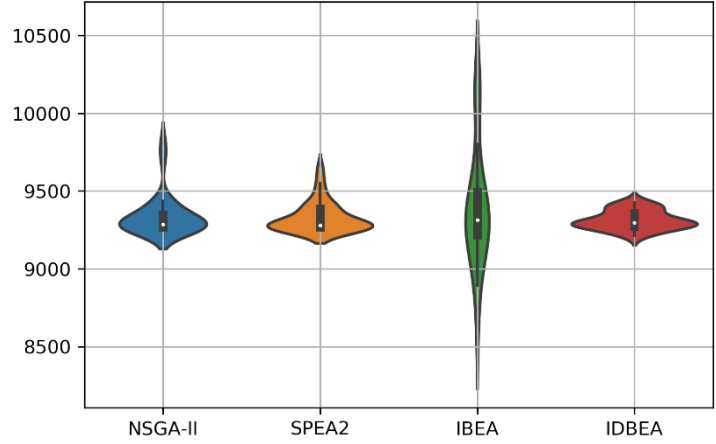

**Fig 8 The cost violin plot of NSGA-II、SPEA2、IBEA and IDBEA for test system 1.**

Fig.8 shows the violin plot of NSGA-II, SPEA2, IBEA, and IDBEA. The violin plot combines the characteristics of the box plot and density graph to display data distribution and probability density. From Fig.8, the distribution of IBEA solutions is too scattered to find the optimal solution, while the solutions of IDBEA near the optimal solution are more intensive and closer to the real optimal solution than other algorithms.

Wilcoxon signed-rank test is a non-parametric statistical test, which is usually used for the performance difference of pair-wise comparison algorithms. In this paper, we propose the following hypothesis:

$H_0$: IDBEA has a significant improvement over the algorithm

In the experiment, all algorithms are executed in the same conditions. We did three independent experiments. Table 2 shows the statistical results of IDBEA and with other three algorithms, while p-value is considered to reject or not $H_0$. In Table 2, p-values are less than the significance level α, so we accept the hypothesis $H_0$: IDBEA has a significant improvement over IBEA, NSGA-II, and SPEA2.

Table 2 Wilcoxon signed-rank test results with significance level α = 0.001.

| Methods | p-value | p-value | p-value |
| --- | --- | --- | --- |
| **IDBEA vs IBEA** | 0.0007 | 0.0001 | 0.0001 |
| **IDBEA vs NAGA-II** | 0.0009 | 0.0004 | 0.0001 |
| **IDBEA vs SPEA2** | 0.0005 | 0.0001 | 0.0001 |

## 5.2 Test system 2

Test system 2 includes a power-only unit (unit1), a heat-only unit (unit5), and three cogeneration units (unit2-4). The system is used to calculate the Pareto set approximation reached with the minimum cost and minimum emission. The power demand and heat demand of the system are chosen to be 300 MW and 150MWth, respectively, meanwhile, the valve-point effects is considered.

As shown in Table 3, the solutions of IBEA and IDBEA are compared in economic dispatch and emission dispatch where IDBEA runs in and 100 solutions are obtained. Besides, the compromise solutions selected in the Pareto sets of NSGA-II, SPEA2, BCS1, GWO, and IBEA are compared. In particular, we choose the best compromise solution when costs and emissions are minimized, as shown in Fig.9, the distribution of the Pareto sets approximation is reached, and the best compromise solution is the solution in the middle of all solutions. The best-found solution of IDBEA in economic dispatch is $13914.8 and the corresponding total emission is 11.7kg, which increased the cost by $58.1 comparing to IBEA and reduced the emission by 0.3 kg. When the emission of pollutants reaches the minimum, the emission of IDBEA is 1.2kg, and the corresponding cost is $17015.3.

Table 3 Results of methods of economic dispatch, emission dispatch and combined economic emission dispatch for system 2 (Pd=300MW, Hd=150MWth).

| Methods | Economic dispatch | | Emission dispatch | | Economic emission dispatch | | | | | |
|---|---|---|---|---|---|---|---|---|---|---|
| | IBEA | IDBEA | IBEA | IDBEA | NSGA-II | SPEA2 | BCS1 | GWO | IBEA | IDBEA |
| P1(MW) | 134.8 | 132.8 | 35 | 35.3 | 87.6 | 90 | 88.2 | 88.5 | 90.3 | 87.1 |
| P2 | 63.2 | 43.8 | 112.4 | 116.9 | 85.2 | 93 | 94.1 | 94.6 | 86.9 | 95.5 |
| P3 | 12.5 | 36.3 | 52.1 | 43.6 | 33.4 | 32.2 | 33.2 | 29 | 38.2 | 17.3 |
| P4 | 90.1 | 87.5 | 101.1 | 104.6 | 94 | 85.1 | 85.1 | 87.9 | 84.7 | 100.2 |
| H2(MWth) | 73 | 59.2 | 93.8 | 90.5 | 73.7 | 72.7 | 72.6 | 54 | 78.3 | 61.4 |
| H3 | 35.5 | 28.4 | 20 | 44.7 | 19.5 | 35.1 | 29.5 | 39.9 | 16.1 | 41 |
| H4 | 9.5 | 3.7 | 0.08 | 0.3 | 9.8 | 10.2 | 12.3 | 21 | 11.2 | 0.2 |
| H5 | 36.7 | 59.8 | 37.8 | 16.7 | 47.3 | 36.8 | 35.7 | 35.2 | 44.8 | 49 |
| Cost($) | 13856.7 | 13914.8 | 17025.9 | 17015.3 | 15188.3 | 15239.2 | 15286.3 | 15243.7 | 15193.5 | 15182 |
| Emission(kg) | 12 | 11.7 | 1.3 | 1.2 | 5.3 | 5.6 | 5.4 | 5.4 | 5.6 | 5.2 |
| Time(s) | 2.8 | 2.3 | 2.8 | 2.3 | 1.9 | 2.0 | 2.3 | 2.5 | 2.8 | 2.3 |

The best compromise points of NSGA-II, SPEA2, BCS1, GWO, and IBEA are selected in Table 3 to compare with IDBEA. The cost and emission of IDBEA are $15182 and 5.2kg. Compared with NSGA-II, SPEA2, BCS1, GWO, and IBEA, the cost of IDBEA is decreased by $6.3, $57.2, $104.3, $61.7 and $11.5, while the emission of IDBEA is decreased by 0.1kg, 0.4kg, 0.2kg, 0.2kg and 0.4kg, respectively. From Table3, the best-found solution obtained by IDBEA reduces the emission and cost more effectively than the other algorithms.

To verify the diversity of solutions, the Pareto Optimal Fronts (POF) obtained by NSGA-II, IBEA, and IDBEA are shown in Fig.9. Each algorithm runs a time, and 100 solutions are generated as the POF:

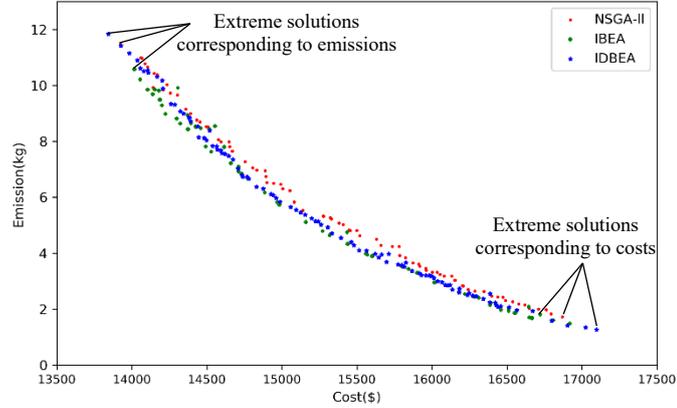

**Fig 9 Distribution of POFs of IBEA、NSGA-II and IDBEA in test system 2.**

As shown in Fig.9, the extreme solutions corresponding to emissions and costs are marked out. The diversity and distribution of IDBEA are better than those of IBEA and NSGA-II. It can be proved that the solutions of IDBEA have better diversity and convergence than the other algorithms. There are several ways to evaluate the multi-objective algorithm, a few measure indicators are used in the literature [32]. In order to compare IDBEA with the results in the literature more convenient, in this paper, two measurement methods are chosen to evaluate the convergence and distribution of solutions: Hypervolume (HV) and Spread($\triangle$). As we know, Hypervolume is a comprehensive indicator, while Spread considers only the diversity of solutions. Specific measurements are shown in Table 4:

Table 4 Quality of the solutions obtained by SPEA2, NSGA-II, IBEA, and IDBEA for test system 2.

| Criteria | SPEA2 | NSGA-II | IBEA | IDBEA |
| --- | --- | --- | --- | --- |
| **HV** | 0.61 | 0.60 | 0.62 | 0.63 |
| $\triangle$ | 0.91 | 0.65 | 0.98 | 0.51 |

As we can learn from section 4.1, the Pareto approximate solution set is closer to the Pareto real solution set, when the hypervolume is greater than another. Spread is different, the more uniform the solution distribution on the Pareto approximate solution set, the closer the value of propagation is to zero. In Table 4, IDBEA has the largest HV value and the smallest Spread value.

In this problem, we will also perform a Wilcoxon signed-rank test. In the experiment, we conducted ten independent experiments, using HV and Spread indicator values for data analysis. The results are shown in Table 5.

Table 5 Wilcoxon signed-rank test results.

| Methods | p-value (HV) | p-value (Spread) |
|---|---|---|
| **IDBEA vs IBEA** | 0.5 | 0.005 |
| **IDBEA vs NAGA-II** | 0.5 | 0.005 |
| **IDBEA vs SPEA2** | 0.4 | 0.005 |

Due to the limited range of quality indicators, the p-values of the HV indicator are less than the significance level 0.5. And the p-values of Spread are less than the significance level 0.01, so IDBEA has a significant improvement over IBEA, NSGA-II, and SPEA2.

In literature [33], the attainment function is used to assess the performance of the algorithm, it's a great job. In literature [34], some tools are provided to describe the probabilistic distribution of the solutions obtained by the algorithm in the objective space. In this paper, the empirical attainment function (EAF) is used to describe the distribution of the solution quality. The best, median and worst attainment surfaces for the solutions of NSGA-II, IBEA, and IDBEA are described in Fig.10.

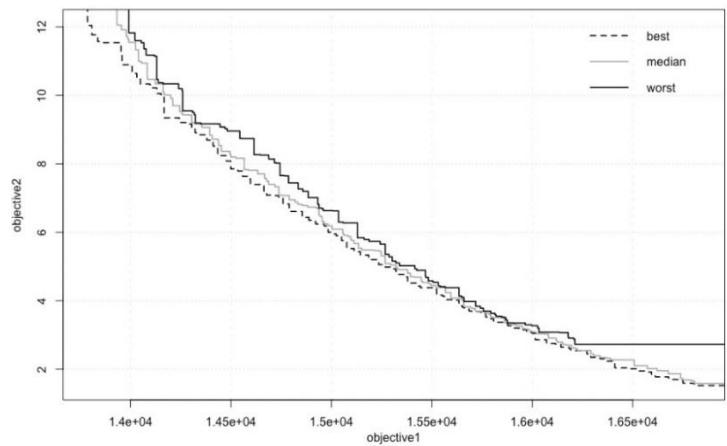

**Fig 10(a) Best, median and worst attainment surfaces of NSGA-II.**

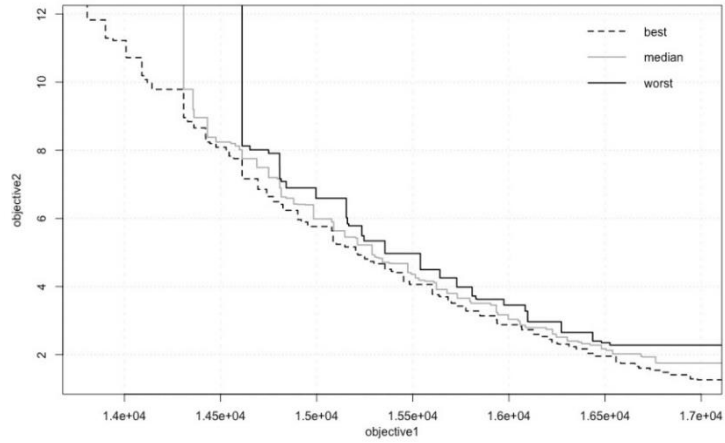

**Fig 10(b) Best, median and worst attainment surfaces of IBEA.**

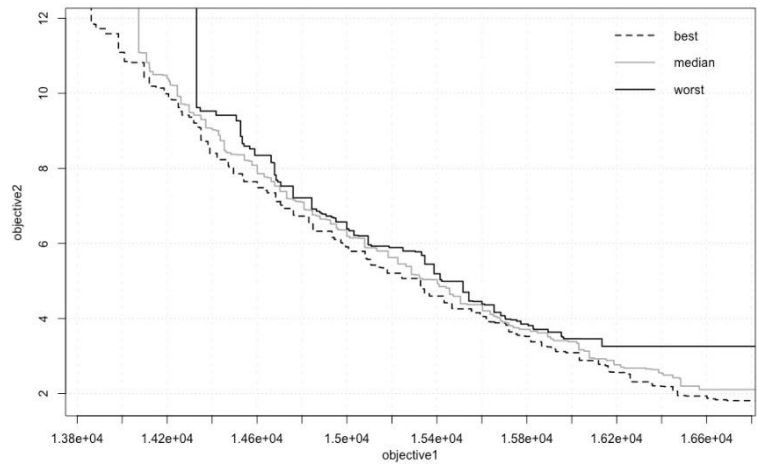

**Fig 10(c) Best, median and worst attainment surfaces of IDBEA.**

In Fig.10, the 25% (worst), 50% (median), and 75%(best) attainment surfaces of the algorithms are plotted. The 50% attainment surface is the true tradeoff solutions as well as the mean attainment surface. We can see in figures, in the intermediate stage, all the solutions converge to the best attainment surfaces. On the left upper side of the figures, IDBEA is worse than NSGA-II while is better than IBEA, in the meanwhile, the region of IDBEA is wider than NSGA-II. In summary, diversity and convergence of IDBEA are better than the other algorithms.

### 5.3    Test system 3

Test system 3 consists of 7 units, including four power-only units (unit1-4), two cogeneration units (unit5-6) and a heat-only unit (unit7). This system considers the valve-point effects and the transmission loss between networks. The parameters and loss factors are detailed in the Appendix. The power demand and heat demand of the system are chosen to be 600MW and 150MWth, respectively.

Table 6 Results of methods of economic dispatch, emission dispatch and combined economic emission dispatch for system 3 (Pd=600MW, Hd=150MWth).

| Methods | Economic dispatch | | Emission dispatch | | Economic emission dispatch | | | | | |
|---|---|---|---|---|---|---|---|---|---|---|
| | IBEA | IDBEA | IBEA | IDBEA | NSGA-II | SPEA2 | BCS1 | GWO | IBEA | IDBEA |
| P1(MW) | 68.6 | 65.7 | 42.1 | 43.4 | 46 | 68.3 | 46 | 54.1 | 61.6 | 64.5 |
| P2 | 92.8 | 97.3 | 59 | 38.2 | 94.4 | 90.9 | 94.4 | 88.2 | 100.4 | 95.8 |
| P3 | 113.5 | 112.4 | 63.2 | 63.9 | 110.4 | 106.5 | 110.4 | 117.6 | 103.8 | 95.5 |
| P4 | 206.2 | 209.2 | 79.8 | 98.8 | 123.3 | 120.2 | 123.3 | 123.1 | 119.3 | 122 |
| P5 | 84.6 | 83.3 | 246.3 | 246.1 | 193.2 | 165 | 193.2 | 188.4 | 157.9 | 188.6 |
| P6 | 42.3 | 40.3 | 116.1 | 116 | 40 | 57.4 | 40 | 40.5 | 63.2 | 40.2 |
| H5(MWth) | 91.7 | 93.9 | 106.3 | 105.6 | 92.2 | 92 | 92.2 | 92.7 | 88.6 | 92.5 |
| H6 | 55.2 | 59.8 | 52.8 | 47.9 | 71.8 | 61.5 | 71.8 | 101.3 | 70.8 | 57 |
| H7 | 5.6 | 0.2 | 0 | 3.4 | 0.4 | 0.9 | 0.4 | 0.5 | 4.2 | 1.6 |
| Ploss(MW) | 6.7 | 6.7 | 6.4 | 6.4 | 6.2 | 6.1 | 7.1 | 7.1 | 6.1 | 6.1 |
| Cost($) | 10418.7 | 10292.3 | 18769.2 | 18676.5 | 13011.1 | 13001.3 | 12968.5 | 12974.1 | 13029.5 | 12957.2 |
| Emission(kg) | 28.3 | 28.8 | 7.6 | 7.7 | 17.4 | 18 | 17.5 | 18 | 18.1 | 17.3 |
| Time(s) | 2.6 | 2.0 | 2.6 | 2.0 | 1.8 | 2.5 | 2.3 | 2.6 | 2.2 | 2.0 |

Table 6 describes the comparison of the minimum cost, minimum emission of the IDBEA with the IBEA, and the comparison of the best compromise solutions of NSGA-II, SPEA2, BCS1, GWO, and IBEA algorithms. The minimum cost of IDBEA in economic dispatch is $10292.3, which is $126.4 less than IBEA, and the emission is 28.8kg, which is 0.5kg more than IBEA. The minimum emission of IDBEA in emission dispatch is 7.7kg, which is 0.1kg more than IBEA, but the cost is $18676.5, which is $92.7 less than IBEA. In the economic emission dispatch, the compromise solution of IDBEA cost $12957.2 and emission is 17.3kg, which decreased $53.9, $44.1, $11.3, $16.9 and $72.3 than NSGA-II, SPEA2, BCS1, GWO, and IBEA in economic dispatch, respectively. Also, IDBEA reduced 0.1kg, 0.7kg, 0.2kg, 0.7kg and 0.8kg emission than NSGA-II, SPEA2, BCS1, GWO, and IBEA, respectively. From Table 6, the transmission loss of IDBEA on this test system is less than that of NSGA-II, BCS1, and GWO. In addition, the transmission loss of IDBEA is equal to IBEA and SPEA2.

To observe the diversity of solutions more intuitively, the Pareto sets approximation reached of NSGA-II, IBEA and IDBEA are given, which are obtained by running the algorithm once. the POF as shown in Fig.11:

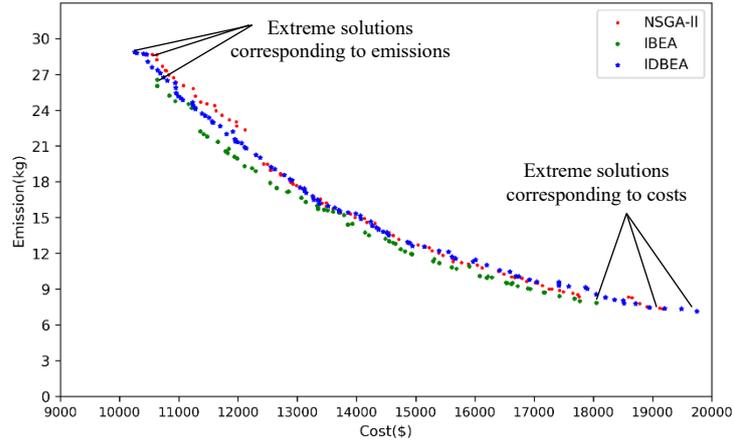

**Fig 11 Distribution of POFs of IBEA、NSGA-II and IDBEA in test system 3.**

Fig.11 depicts the corresponding extreme values of the cost and emission of the solutions of NSGA-II, IBEA, and IDBEA. It indicates that the distribution of the corresponding extremes of the cost and emission of IDBEA is broader than that of NSGA-II and IBEA, while the best-found solutions obtained are more diverse. Hypervolume and Spread measurements of each algorithm are shown in Table 7.

Table 7 Quality of the solutions obtained by SPEA2, NSGA-II, IBEA, and IDBEA for test system 3.

| Criteria | SPEA2 | NSGA-II | IBEA | IDBEA |
|---|---|---|---|---|
| **HV** | 0.63 | 0.64 | 0.63 | 0.65 |
| **Δ** | 0.49 | 0.50 | 0.88 | 0.46 |

In Table 7, the Hypervolume of IDBEA has the maximum value and Spread has the minimum value among the four algorithms.

The Wilcoxon signed-rank test results are shown in Table 8. The p-values of HV are less than the significance level 0.5, and the p-values of Spread are less than the significance level 0.01. So IDBEA has a significant improvement over IBEA, NSGA-II, and SPEA2.

Table 8 Wilcoxon signed-rank test results.

| Methods | p-value (HV) | p-value (Spread) |
|---|---|---|
| **IDBEA vs IBEA** | 0.3 | 0.005 |
| **IDBEA vs NAGA-II** | 0.03 | 0.006 |
| **IDBEA vs SPEA2** | 0.02 | 0.005 |

In Fig.12, the plots of the empirical attainment function are shown, and the best, median and worst attainment surfaces for the solutions are described.

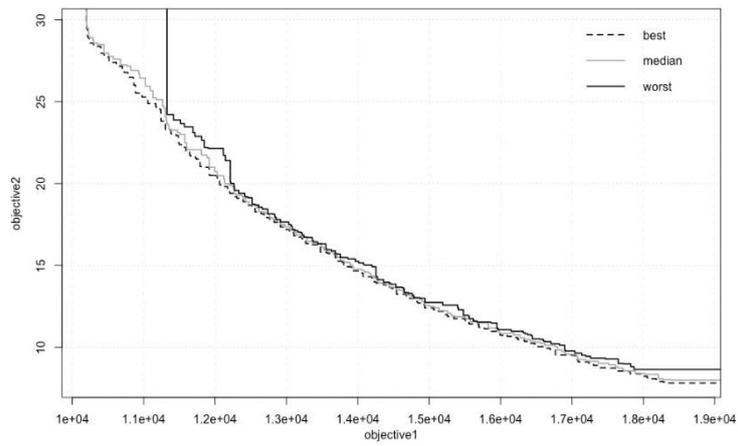

**Fig 12(a) Best, median and worst attainment surfaces of NSGA-II.**

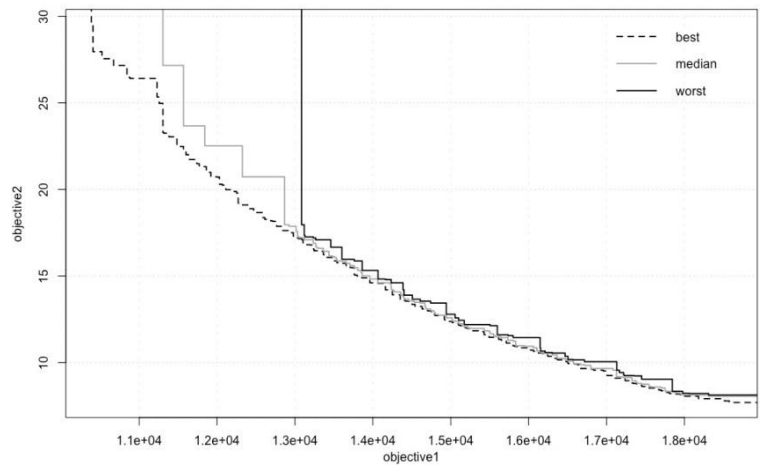

**Fig 12(b) Best, median and worst attainment surfaces of IBEA.**

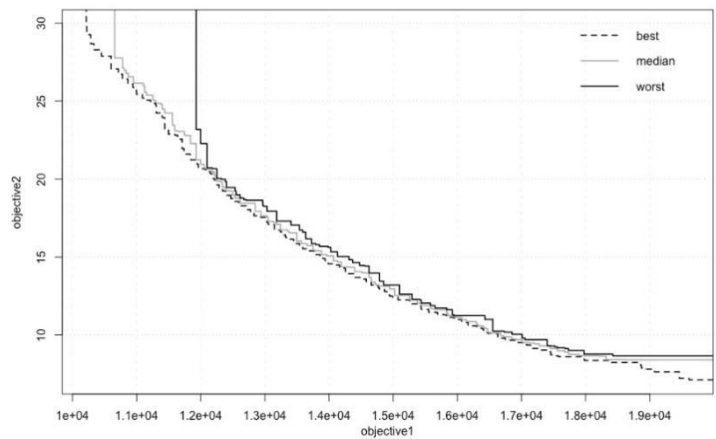

**Fig 12(c) Best, median and worst attainment surfaces of IDBEA.**

In Fig.12, IDBEA shows better convergence than IBEA in the left upper side of the figures, and the distribution of IDBEA is wider than IBEA.

From the experiments above, we can find that IDBEA shows not only stable characteristics and the efficiency in the CHPED problem but also the convergence to the best solutions in CHPEED problems. Compared with the Pareto set approximation reached of IBEA, IDBEA shows more diversity. To conclude, IDBEA has more advantages of convergence and diversity in energy saving and emission reduction than the other existing algorithms, and more in line with the current social demand for environmental protection. Thus, IDBEA can follow the theme of social development closely and provide more effective solutions to meet social needs. IDBEA has strong robustness and extensiveness and can be effectively used to solve both single-objective and multi-objective problems of CHP.

# 6  Conclusion

To meet the needs of energy saving and environmental protection in CHPEED problems, an Indicator & crowding Distance-Based Evolutionary Algorithm (IDBEA) is put forward with consideration of the valve-point effects and transmission loss. To verify the universality and effectiveness of IDBEA, three typical standard test systems are chosen to compare the convergence and diversity of the best solutions with the other reported algorithms. For the CHPED problem, the best-found solution of IDBEA is reduced by $4 compared to the optimal cost of the other algorithms in the first test system. For the CHPEED problem, the optimal compromise solution of IDBEA is reduced by $104.3, 0.4kg and $72.3, 0.8kg comparing to the best-found solutions in the other algorithms in two systems, respectively. For environmental protection, IDBEA focuses more on emission reduction, which is of great practical significance. We still need to combine evolutionary multi-objective optimization theory, although the Hypervolume indicator has certain benefits, there may still be deficiencies for different problems, a better indicator is considered for evaluation. In future work, we will consider how to select the best compromise solution automatically to meet the preferences of decision makers in different situations. In addition, in the process of searching the best solution, we intend to adopt more efficient methods, such as parallel computing, to improve the practicability of the algorithm. When we use large-scale software in our future work, we will use new technologies in literature [35] [36] to speed up the understanding of the software and conduct large-scale experiments better.

# 7  Acknowledgement

The work is supported by the National Natural Science Foundation of China (Grant No. 61876138), the Special Fund for Key Discipline Construction of General Institutions of Higher Learning from Shaanxi Province, the Industrial Research Project of Xi'an (2019218114GXRC017CG018-GXYD17.10), and the Innovation Fund of Xi'an University of Posts and Telecommunications (CXJJLY2018053).

# 8   References

[1] Jayakumar N, Subramanian S, Ganesan S, et al, Grey wolf optimization for combined heat and power dispatch with cogeneration systems, Int J Electr Power Energy Syst. 74(2016) 252-264. https://doi.org/10.1016/j.ijepes.2015.07.031.

[2] Johan Karlsson, Lena Brunzell, G. Venkatesh, Material-flow analysis, energy analysis, and partial environmental-LCA of a district-heating combined heat and power plant in Sweden, Energy. 144(2018) 31-40. https://doi.org/10.1016/j.energy.2017.11.159.

[3] Himanshu Anand, Nitin Narang, J.S. Dhillon, Multi-objective combined heat and power unit commitment using particle swarm optimization, Energy. 172(2019) 794-807. https://doi.org/10.1016/j.energy.2019.01.155.

[4] Basu M, Group search optimization for combined heat and power economic dispatch, Int J Electr Power Energy Syst. 78(2016) 138-147. https://doi.org/10.1016/j.ijepes.2015.11.069.

[5] Nazari-Heris M, Mohammadi-Ivatloo B, Gharehpetian G B, A comprehensive review of heuristic optimization algorithms for optimal combined heat and power dispatch from economic and environmental perspectives, Renew Sust Energ Rev. 81(2)(2018) 2128-2143. https://doi.org/10.1016/j.rser.2017.06.024.

[6] Ghorbani, Naser, Combined heat and power economic dispatch using exchange market algorithm, Int J Electr Power Energy Syst. 82(2016) 58-66. https://doi.org/10.1016/j.ijepes.2016.03.004.

[7] Yang L, Jinlong W, Dongbo Z, et al, A Two-Stage Approach for Combined Heat and Power Economic Emission Dispatch: Combining Multi-Objective Optimization with Integrated Decision Making, Energy. 462(2018) 237– 254. https://doi.org/10.1016/j.energy.2018.07.200.

[8] Niknam T, Azizipanah-Abarghooee R, Roosta A, et al, A new multi-objective reserve constrained combined heat and power dynamic economic emission dispatch, Energy. 42(1)( 2012) 530-545. https://doi.org/10.1016/j.ijepes.2016.03.004.

[9] Guo T, Henwood MI, van Ooijen M, An algorithm for combined heat and power economic dispatch, IEEE Trans Power Syst. 11(4)( 1996) 1778–1784. https://doi.org/10.1109/59.544642.

[10] Rooijers F J, Van Amerongen R A M, Static economic dispatch for co-generation systems, IEEE T Power Syst. 9(3)(1994) 1392-1398. https://doi.org/10.1109/59.336125.

[11] Chang C S, Fu W, Stochastic multiobjective generation dispatch of combined heat and power systems, IEE Proceedings - Generation, Transmission and Distribution. 145(5)(1998) 583-591. https://digital-library.theiet.org/content/journals/10.1049/ip-gtd_19981997.

[12] Nguyen T T, Vo D N, Dinh B H, Cuckoo search algorithm for combined heat and power economic dispatch, Int J Electr Power Energy Syst. 81(2016) 204-214. https://doi.org/10.1016/j.ijepes.2016.02.026.

[13] Nitin Narang, Era Sharma, J.S. Dhillon, Combined heat and power economic dispatch using

integrated civilized swarm optimization and Powell's pattern search method, Appl Soft Comput. 52(2017) 190-202. https://doi.org/10.1016/j.asoc.2016.12.046.

[14] Murugan R, Mohan M R, Christober A R C, et al, Hybridizing Bat Algorithm with Artificial Bee Colony for Combined Heat and Power Economic Dispatch, Appl Soft Comput. 72(2018) 189-217. https://doi.org/10.1016/j.asoc.2018.06.034.

[15] Jiaze S, Yang L, Social cognitive optimization with tent map for combined heat and power economic dispatch, Int Trans Electr Energ Syst. 29(1)(2019) e2660. https://doi.org/10.1002/etep.2660.

[16] Qiang Z, Dexuan Z, Na D, et al, An adaptive differential evolutionary algorithm incorporating multiple mutation strategies for the economic load dispatch problem, Appl Soft Comput, 78(2019) 641-669. https://doi.org/10.1016/j.asoc.2019.03.019.

[17] Yakupov I, Buzdalov M, Improved incremental non-dominated sorting for steady-state evolutionary multiobjective optimization, GECCO'17.ACM. (2017)649-656. https://doi.org/10.1145/3071178.3071307.

[18] Zhengping L, Xuyong W, Qiuzhen L, et al, A novel multi-objective co-evolutionary algorithm based on decomposition approach, Appl Soft Comput. 73(2018) 50-66. https://doi.org/10.1016/j.asoc.2018.08.020.

[19] Yongkuan Y, Jianchang L, Shubin T, et al, A multi-objective differential evolutionary algorithm for constrained multi-objective optimization problems with low feasible ratio, Appl Soft Comput. 80(2019) 42-56. https://doi.org/10.1016/j.asoc.2019.02.041.

[20] Deb K, Agrawal S, Pratap A, et al. A fast elitist non-dominated sorting genetic algorithm for multi-objective optimization: NSGA-II. Evol Comput, 1917(2000) 849-858. https://doi.org/10.1007/3-540-45356-3_83.

[21] Basu M, Combined heat and power economic emission dispatch using nondominated sorting genetic algorithm-II, Int J Electr Power Energy Syst. 53(1)(2013) 135-141. https://doi.org/10.1016/j.ijepes.2013.04.014.

[22] Wang L, Singh C, Stochastic combined heat and power dispatch based on multi-objective particle swarm optimization, Electr Power Energy Syst. 30(3)(2008) 226–234. https://doi.org/10.1016/j.ijepes.2007.08.002.

[23] Yousef ali Shaabani, Ali Reza Seifi, Masoud Joker Kouhanjani, Stochastic multi-objective optimization of combined heat and power economic/emission dispatch, Energy. 141(2017) 1892-1904. https://doi.org/10.1016/j.energy.2017.11.124.

[24] Shi B, Yan L X, Wu W, Multi-objective optimization for combined heat and power economic dispatch with power transmission loss and emission reduction, Energy. 56(2013) 135-143. https://doi.org/10.1016/j.energy.2013.04.066.

[25] Zhang Q, Li H, MOEA/D: A Multiobjective Evolutionary Algorithm Based on Decomposition, IEEE T Evolut Comput. 11(6)( 2008) 712-731. https://doi.org/10.1109/TMAG.2011.2174348.

[26] Zitzler E, Simon Künzli, Indicator-based selection in multiobjective search, Parallel Problem

Solving from Nature - PPSN VIII. PPSN 2004. Lecture Notes in Computer Science. 3242(2004) 832-842. https://doi.org/10.1007/978-3-540-30217-9_84.

[27] Ehsan Naderi, Mahdi Pourakbari-Kasmaei, Hamdi Abdi, An efficient particle swarm optimization algorithm to solve optimal power flow problem integrated with FACTS devices, Appl Soft Comput. 80(2019) 243-262. https://doi.org/10.1016/j.asoc.2019.04.012.

[28] Zitzler E, Thiele L, Multiobjective evolutionary algorithms: a comparative case study and the strength Pareto approach, Evol. Compu. 3(4)(1999) 257-271. https://doi.org/10.1109/4235.797969.

[29] Yinxing X, Jinghui Z, Tian Huat T, et al, IBED: Combining IBEA and DE for optimal feature selection in software product line engineering, Appl Soft Comput. 49(2016) 1215-1231. https://doi.org/10.1016/j.asoc.2016.07.040.

[30] Wagner T, Beume N, Naujoks B, Pareto-, aggregation-, and indicator-based methods in many-objective optimization, EMO. 4403(2007) 1973 – 2019. https://doi.org/10.1007/978-3-540-70928-2_56.

[31] Niknam T, Azizipanah-Abarghooee R, Roosta A, et al, A new multi-objective reserve constrained combined heat and power dynamic economic emission dispatch, Energy. 42(1)( 2012) 530-545. https://doi.org/10.1016/j.ijepes.2016.03.004.

[32] N. Riquelme, C. Von Lücken, B. Baran, Performance metrics in multi-objective optimization, 2015 Latin American Computing Conference (CLEI), Arequipa. (2015) 1-11. https://doi.org/10.1109/CLEI.2015.7360024.

[33] Grunert da Fonseca V., Fonseca C.M., Hall A.O, Inferential Performance Assessment of Stochastic Optimisers and the Attainment Function, (eds) Evolutionary Multi-Criterion Optimization. EMO 2001. Lecture Notes in Computer Science. 1993(2001) 213-225. https://doi.org/10.1007/3-540-44719-9_15.

[34] López-Ibáñez M., Paquete L., Stützle T, Exploratory Analysis of Stochastic Local Search Algorithms in Biobjective Optimization, (eds) Experimental Methods for the Analysis of Optimization Algorithms. (2010) 209-222. https://doi.org/10.1007/978-3-642-02538-9_9.

[35] Weifeng P, Bing L, Jing L, et al, Analyzing the structure of Java software systems by weighted k-core decomposition. Future Gener Comp Sy. 83(2018) 431-444. https://doi.org/10.1016/j.future.2017.09.039.

[36] Weifeng P, Beibei S, Kangshun L, et al, Identifying Key Classes in Object-Oriented Software using Generalized k-Core Decomposition. Future Gener Comp Sy. 81(2018) 188-202. https://doi.org/10.1016/j.future.2017.10.006.

# 9 Appendix A.

### A.1.1 Cost and emission functions of each unit of test system 2

**(a) Power-only units**

$$C_{t1}(P_1) = 254.8863 + 7.6997P_1 + 0.00172P_1^2 + 0.000115P_1^3 \; \$$$

$$35 \leq P_1 \leq 135 \; \text{MW}$$

$$E_{t1}(P_1) = 10^{-4} \times (4.091 - 5.554P_1 + 6.490P_1^2) + 2 \times 10^{-4} \times exp(0.02857P^1) \; kg$$

**(b) Cogeneration units**

$$C_{c2}(P_2, H_2) = 1250 + 36P_2 + 0.0435P_2^2 + 0.6H_2 + 0.027H_2^2 + 0.011P_2H_2 \; \$$$

$$E_{c2}(P_2, H_2) = 0.00165P_2 \; kg$$

$$C_{c3}(P_3, H_3) = 2650 + 34.5P_3 + 0.1035P_3^2 + 2.203H_3 + 0.025H_3^2 + 0.051P_3H_3 \; \$$$

$$E_{c3}(P_3, H_3) = 0.0022P_3 \; kg$$

$$C_{c4}(P_4, H_4) = 1565 + 20P_4 + 0.072P_4^2 + 2.3H_4 + 0.02H_4^2 + 0.04P_4H_4 \; \$$$

$$E_{c4}(P_4, H_4) = 0.0011P_4 \; kg$$

**(c) Heat-only unit**

$$C_{h5}(H_5) = 950 + 2.0109H_5 + 0.038H_5^2 \; \$$$

$$0 \leq H_5 \leq 60 \; \text{MWth}$$

$$E_{h5}(H_5) = 0.0017H_5 \; kg$$

**A.1.2 Heat-power FOR of cogeneration units**

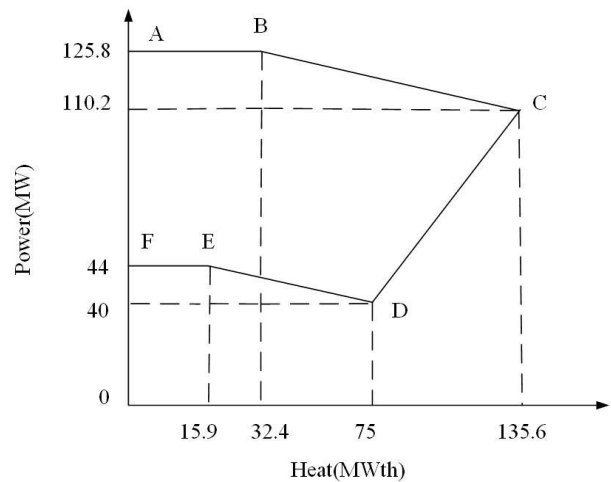

**Fig 13 FOR of cogeneration unit 1 in test system 2.**

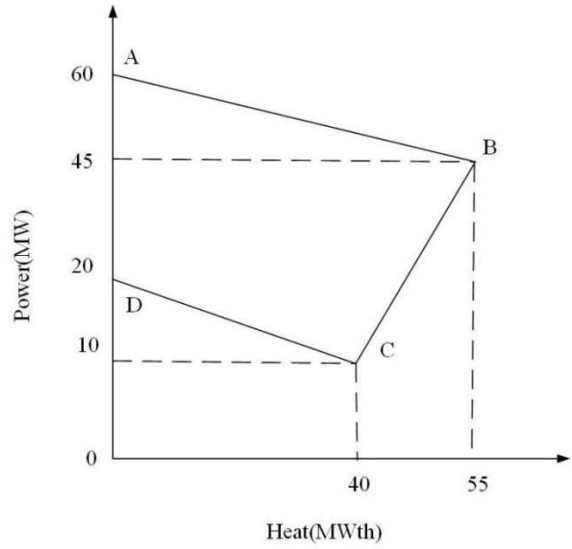

**Fig 14 FOR of cogeneration unit 2 in test system 2.**

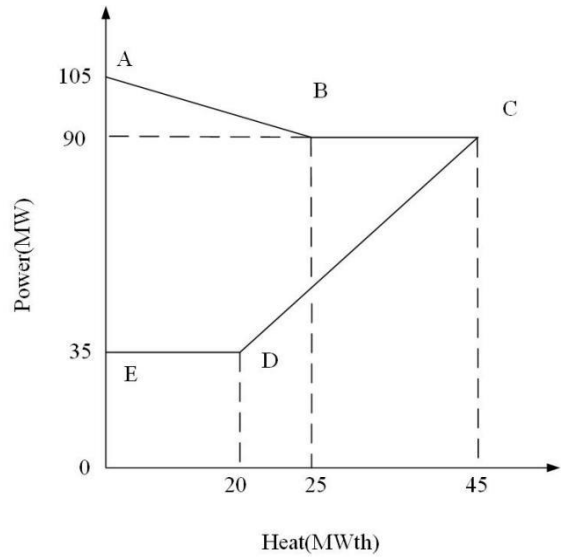

**Fig 15 FOR of cogeneration unit 3 in test system 2.**

## A.2.1 Cost and emission functions of each unit of test system 3

### (a) Power-only units

$$C_{t1}(P_1) = 25 + 2P_1 + 0.008P_1^2 + \left|100\sin\{0.042(P_1^{min} - P_1)\}\right| \; \$$$

$$10 \leq P_1 \leq 75 \; \text{MW}$$

$$E_{t1}(P_1) = 10^{-4} \times (4.091 - 5.554P_1 + 6.490P_1^2) + 2 \times 10^{-4} \times exp(0.02857P_1) \; \text{kg}$$

$$C_{t2}(P_2) = 60 + 1.8P_2 + 0.003P_2^2 + \left|140\sin\{0.04(P_2^{min} - P_2)\}\right| \; \$$$

$$20 \leq P_2 \leq 125 \; \text{MW}$$

$$E_{t2}(P_2) = 10^{-4} \times (2.543 - 6.047P_2 + 5.638P_2^2) + 5 \times 10^{-4} \times exp(0.03333P_2) \; \text{kg}$$

$$C_{t3}(P_3) = 100 + 2.1P_3 + 0.0012P_3^2 + \left|160\sin\{0.038(P_3^{min} - P_3)\}\right| \; \$$$

$$30 \leq P_3 \leq 175 \; \text{MW}$$

$$E_{t3}(P_3) = 10^{-4} \times (4.258 - 5.094P_3 + 4.586P_3^2) + 1 \times 10^{-6} \times exp(0.08P_3) \; \text{kg}$$

$$C_{t4}(P_4) = 120 + 2P_4 + 0.001P_4^2 + \left|180\sin\{0.037(P_4^{min} - P_4)\}\right| \; \$$$

$$40 \leq P_4 \leq 250 \; \text{MW}$$

$$E_{t4}(P_4) = 10^{-4} \times (5.326 - 3.550P_4 + 3.370P_4^2) + 2 \times 10^{-3} \times exp(0.02P_4) \; \text{kg}$$

**(b) Cogeneration units**

$$C_{c5}(P_5, H_5) = 2650 + 14.5P_5 + 0.0345P_5^2 + 4.2H_5 + 0.03H_5^2 + 0.031P_5H_5 \; \$$$

$$E_{c5}(P_5, H_5) = 0.00165P_5 \; kg$$

$$C_{c6}(P_6, H_6) = 1250 + 36P_6 + 0.0435P_6^2 + 0.6H_6 + 0.027H_6^2 + 0.011P_6H_6 \; \$$$

$$E_{c6}(P_6, H_6) = 0.00165P_6 \; kg$$

**(c) Heat-only unit**

$$C_{h7}(H_7) = 950 + 2.0109H_7 + 0.038H_7^2 \; \$$$

$$0 \leq H_7 \leq 2695.2 \; \text{MWth}$$

$$E_{h7}(H_7) = 0.0018H_7 \; kg$$

**(d) Network loss coefficients**

$$B = \begin{bmatrix} 49 & 14 & 15 & 15 & 20 & 25 \\ 14 & 45 & 16 & 20 & 18 & 19 \\ 15 & 16 & 39 & 10 & 12 & 15 \\ 15 & 20 & 10 & 40 & 14 & 11 \\ 20 & 18 & 12 & 14 & 35 & 17 \\ 25 & 19 & 15 & 11 & 17 & 39 \end{bmatrix} \times 10^{-6}$$

$$B_0 = [-0.3908 \quad -0.1297 \quad 0.7047 \quad 0.0591 \quad 0.2161 \quad -0.6635] \times 10^{-3}$$

$$B_{00} = 0.056$$

## A.2.2 Heat-power FOR of cogeneration units

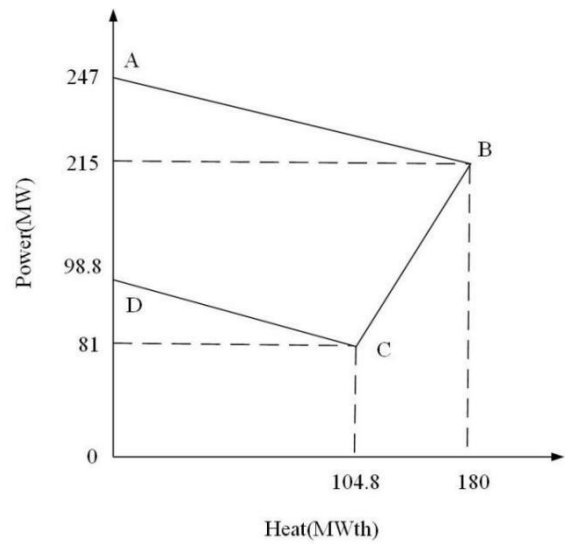

**Fig 16 FOR of cogeneration unit 1 in test system 3.**

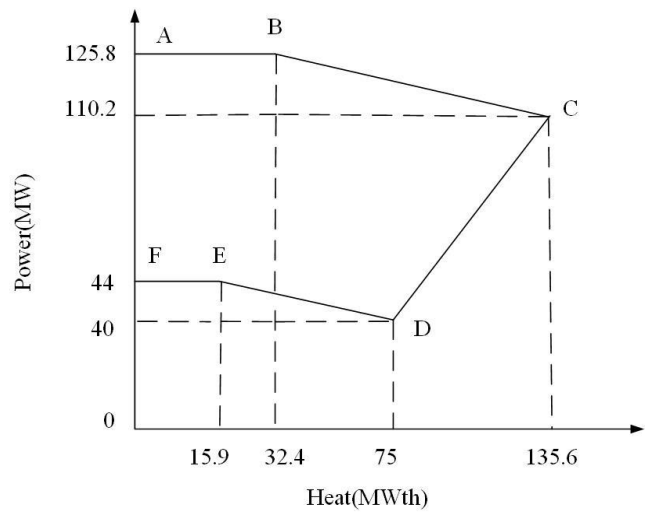

**Fig 17 FOR of cogeneration unit 2 in test system 3.**